\documentclass[preprint]{revtex4}
\usepackage{}
\usepackage{epsfig,latexsym,amssymb}
\usepackage{latexsym}
\usepackage{amssymb}
\usepackage{amsfonts}
\usepackage{comment}
\usepackage{graphicx}
\usepackage{verbatim}

\newcommand{\be}{\begin{equation}}
\newcommand{\ee}{\end{equation}}
\newcommand{\bea}{\begin{eqnarray}}
\newcommand{\eea}{\end{eqnarray}}

\begin{document}

\title{Hemispherical Power Asymmetry of the Cosmic Microwave Background from a Remnant of a pre-Inflationary Topological Defect}

\author{Qiaoli Yang$^{1,}$\footnote{corresponding author\\qiaoli\_yang@hotmail.com}, Yunqi Liu$^{2,}$\footnote{liuyunqi@hust.edu.cn}, and Haoran Di$^{2,}$\footnote{haoran\_di@yahoo.com}}
\affiliation{$^1$Physics Department, Jinan University, Guangzhou 510632, China\\$^2$School of Physics, Huazhong University of Science and Technology, Wuhan 430074, China}

\begin{abstract}
Observations indicate that large-scale anomalies exist in the fluctuations of the cosmic microwave background. In these anomalies, the hemispherical power amplitude asymmetry has a correlation length comparable to that of the observable universe. We propose that a topological defect created by spontaneous breaking of the U(1) symmetry prior to inflation generated an initial phase variation, $\delta \theta$, across the observable region of the universe. The amplitude of this phase fluctuation is protected by topology if the defect is inside the horizon, and is frozen by causality if the defect exits the horizon. After inflation, the phase-corresponding boson field started to oscillate, when the Hubble rate decreased to a level comparable to the mass of the boson field. The energy density of the newly created boson particles varied across the observable universe. The bosons subsequently decayed into radiation prior to the BBN epoch, and the resulting fluctuations in the energy density produced the observed power asymmetry. This scenario predicts a scale-dependent modulation amplitude power asymmetry and in addition, as topological defects created by phase transitions are a very general phenomenon, the observed hemispherical asymmetry may be seen as an evidence for the cosmological inflation.
\end{abstract}

\date{\today}
\maketitle

\section{Introduction}
%{\itshape Introduction:}
Observations of the cosmic microwave background (CMB) are consistent with the inflation paradigm, especially at small scales. In this paradigm \cite{Guth:1980zm,Albrecht:1982wi,Linde:1981mu}, quantum fluctuations seed cosmic structures, and the predictions of this theory agree well with the observations \cite{Komatsu:2010fb,Ade:2013zuv}. However, measurements from the Wilkinson Microwave Anisotropy (WMAP) and the PLANCK mission indicate that there are large-scale anomalies, in which the hemispherical power asymmetry has a correlation length comparable to the observable universe \cite{Eriksen:2003db,Hansen:2004vq,Land:2005ad,Eriksen:2007pc,Hansen:2008ym,Hoftuft:2009rq,Paci:2013gs,Akrami:2014eta,Ade:2015hxq,Ade:2013nlj}.

The hemispherical power asymmetry can be written as:
\be
\Delta T(\hat n)=(1+A\hat p\cdot\hat n)\Delta T_{iso}(\hat n)~~,
\ee
where $\hat n$ is a unit vector of the observed direction, $\hat p$ is the preferred direction of the asymmetry and $A=0.072\pm0.022$ as determined from the WMAP and the PLANCK mission \cite{Donoghue:2004gu,Hanson:2009gu,Bennett:2010jb} on large scales. On small scales ($l=601-2048$), $A<0.0045$ which suggests a strong scale dependent property \cite{Hirata:2009ar,Flender:2013jja,Adhikari:2014mua,Aiola:2015rqa}. There are proposals that may explain the asymmetry such as \cite{Turner:1991dn,GarciaBellido:1997te,Erickcek:2009at,Kanno:2013ohv,Lyth:2013vha,DAmico:2013hur,Kohri:2013kqa,Assadullahi:2014pya,Jazayeri:2014nya,Lyth:2001nq,Moroi:2001ct,Enqvist:2001zp,Malik:2002jb,Langlois:2004nn,Linde:2005yw,Erickcek:2008sm,Cai:2013gma,Cai:2015xba,Byrnes:2015asa,Byrnes:2015dub,Kobayashi:2015qma,Ringeval:2015ywa,Ashoorioon:2015pia,Byrnes:2016uqw} etc.

The standard inflation theory proposes that the universe experienced a period of exponential expansion, and therefore the observable universe was once a region within a horizon, with an causal contact established. The spatial curvature and exotic-relics such as the magnetic-monopoles etc. were diminished by inflation. Thus the universe today is smooth, flat and homogeneous. In addition, quantum fluctuations created the primordial density fluctuations which caused the current cosmic structures. The predictions of the theory agree well with the observations, but the large scale anomalies in the CMB suggest that refinements are needed.

In this paper, we propose that a topological defect, produced by spontaneous symmetry breaking prior to inflation, produced the observed power asymmetry.

The creation of topological defects by phase transitions is a common phenomenon in field theory and in condensed matter physics \cite{Vilenkin:1982ks,Harari:1987ht,Sakharov:1994id,Hindmarsh:1994re,Sakharov:1996xg,Khlopov:1999tm,Sikivie:2006ni,Khlopov:2008qy,Zhang:2015bga}. Let us consider here a simple complex scalar field with a U(1) symmetry potential plus a mass term. The U(1) symmetry of the complex scalar broke spontaneously prior to inflation, and string-like topological defects, with a typical phase variation $\Delta\theta=2\pi$ around the defects, were created. Part of the variation $\delta \theta$ was located in the region that evolved to become the current observable universe (see Fig.1). As the $\delta \theta$ variation is protected by topology if the defect is within the horizon and is frozen by causality if the defect exits the horizon, the variation remains a constant through inflation. After inflation, and when the Hubble rate of the universe decreased to comparable to the mass of the phase corresponding boson field, the field rolled toward its minimum and started to oscillate. Thus a condensate of boson particles with a primordial density variation was created. The bosons later decayed into radiations and the variation of the boson particles generated the observed hemispherical power asymmetry.

In this scenario, as in the curvaton case, one additional quantum field besides the inflaton played an important role in the formation of CMB power spectrum. However, the new introduced U(1) symmetry of the field plays an essential role in the formation of the power asymmetry. The spontaneous symmetry breaking created topological defect naturally generated the required primordial field amplitude asymmetry. In addition, the resulted power asymmetry is scale-dependent which is preferred by recent cosmological observations.

\begin{figure}
\begin{center}
\includegraphics[width=0.9\textwidth]{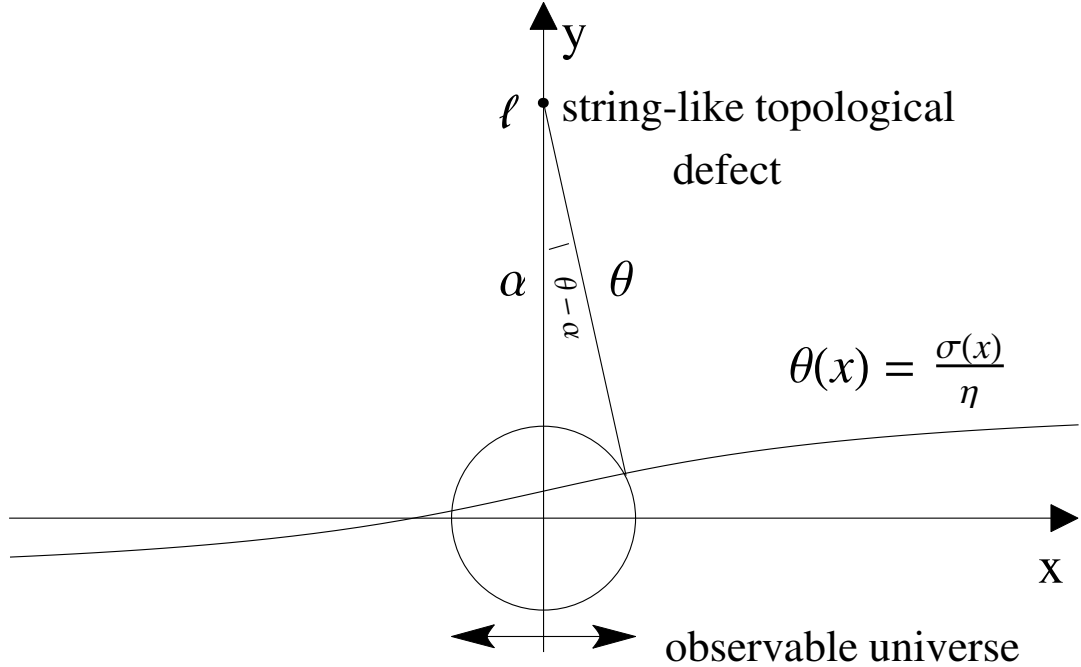}
\caption{The fluctuation created by a string like topological defect. The $\theta$ term corresponds to a boson field with an amplitude $\sigma=\theta\eta$, where $\eta$ is the symmetry breaking scale. As the spontaneous symmetry breaking happens prior to inflation, $\sigma\sim \eta>H_I$. An explicitly symmetry breaking potential gives rise to a mass to the $\sigma$ field. The minimal energy configuration of the boson field is $\theta=0$ thus $\Sigma=\alpha\eta$ corresponds to the homogeneous background value of the field and $\delta \theta=\theta-\alpha$ corresponds to the field amplitude fluctuation.}
\end{center}
\end{figure}

\section{Inflation and Primordial Density Perturbation}
%{\itshape Inflation and Primordial Density Perturbation:}
The single-field inflation theory \cite{Bassett:2005xm} predicts that the energy density of the universe was dominated by the inflaton field $\phi$. When the potential energy of the inflaton dominates over its kinetic energy, the expansion rate of the universe is accelerated, which solves many puzzles in the big bang theory. To achieve this scenario, the potential of the inflaton field $V(\phi)$ satisfies the slow-roll condition: $(m^2_{pl}/16\pi)(V'/V)^2\ll 1$ and $(m_{pl}^2/8\pi)(V''/V)\ll1$, where $m_{pl}=\sqrt G=1.2\times10^{19}$GeV is the Planck mass. The universe experienced a period of exponentially expanse $a(t_{end})/a(t_{begin})\sim e^N$ where $a(t)$ is the scale factor and $N\sim 60$.

The primordial density fluctuation that is responsible for the structure formation is due to the quantum fluctuations of the inflaton. Considering only the linear scalar perturbations, the perturbed Friedmann-Robertson-Walker (FRW) spacetime can be parameterized by the Newtonian potential $\Psi$ and the spatial curvature perturbation $\Phi$ as follows:
\be
ds^2=-(1+2\Psi)dt^2+a^2(1-2\Phi)\delta_{ij}dx^idx^j~~.
\ee
As the anisotropic stress is negligible \cite{Hu:1994uz}, we have $\Phi=\Psi$. The power spectrum of gravitation potential is defined as: $\langle\Phi(\vec k)\Phi(\vec k')\rangle=(2\pi)^3P_{\Phi}(k)\delta^3(\vec k-\vec k')$.

As the inflation paradigm predicts that the power spectrum will be isotropic, and additionally, physical processes occurring after inflation should have a correlation length smaller than the size of the observable universe, the existence of a hemispherical power asymmetry across the observable universe is unexpected. A cosmic structure created by a spontaneous symmetry breaking prior to inflation giving rise to the observed power asymmetry appears to be a reasonable solution. In the following discussions, we propose a scenario in which the observed CMB asymmetry is a remnant from a pre-inflationary string-like topological defect of a boson field.

\section{Pseudo-Goldstone Bosons and String-like Topological Defects}
%{\itshape Pseudo-Goldstone Bosons and String-like Topological Defects:}
We now consider a complex field, $\tau$, with a U(1) symmetry potential plus a potential term, $V$. The relative Lagrangian is:
\be
{\cal L}\supset {1\over 2}\lambda^2(\tau^\dagger\tau-\eta^2)^2+V(\theta)~~,
\ee
where $\lambda$ and $\eta$ are positive parameters and $\theta$ is the phase angle of the complex field $\tau$. The U(1) symmetry is spontaneously broken and $\tau$ acquires a vacuum expectation value,
\be
\langle\tau\rangle=\eta e^{i\theta(x)}~~.
\ee
The $V(\theta)$ term gives rise to a mass $m_{\sigma}$ to the Goldstone boson thus it becomes a Pseudo-Goldstone.

The Lagrangian receives additional contributions from a non-zero temperature therefore when $T>T_c\sim\eta$ \cite{Vilenkin:1982ks}, the minimum of the potential is at $\tau=0$, and the U(1) symmetry is restored. As in this paper, the U(1) symmetry is spontaneously broken prior to inflation, we have: $T_I\sim H_I/(2\pi)$ \cite{Gibbons:1977mu}, and $\eta>H_I/(2\pi)$.

Immediately after the U(1) symmetry is broken, the new created Pseudo-Goldstone bosons satisfy:
\be
(\partial_t^2+{3H}\partial_t+{k^2\over a^2})\sigma(\vec k ,t)=0~~,
\label{horizon}
\ee
where the mass term is negligible during this era. For the modes with a comoving wave length $2\pi (k/a)^{-1}\gg H^{-1}$, the last term of Eq.(\ref{horizon}) is negligible, therefore the solution is $\sigma(\vec k,t)\sim \sigma(\vec k)$ which is often referred as "frozen by causality". For the modes with a comoving wave length $2\pi (k/a)^{-1}\ll H^{-1}$, the modes oscillate with a decreasing amplitude, thus sub-horizon fluctuations are typically diluted due to inflation.

If topological defects appear, the fluctuations can be protected by the topology. Let us consider the cosmic strings created by the breaking of the global U(1) symmetry. After the spontaneous symmetry breaking, the phase angle $\theta$ varies around a closed loop if a defect appears inside the loop. The total change of the phase around a closed loop is $2\pi$ when the system is in the lowest energy state. Using the cylindrical coordinates, the field configuration around a cosmic string can be written as: $\tau=\eta e^{i\theta}$,
for $r>r_{core}$ where $r_{core}\sim \lambda^{-1}\eta^{-1}$ is the radius of the string core. The amplitude of the field is substantially different from $\eta$ only within the string core. Thus the energy per unit length of the string is $\mu\sim 2\pi\eta^2ln(R /r_{core})$ \cite{Vilenkin:1982ks}, where the radius $R$ is the typical distance between adjacent strings.

\section{The Initial Fluctuation}
%{\itshape The Initial Fluctuation:}
We now consider a scenario where a string-like topological defect was created prior to inflation and near the region which evolved to the observable universe. The initial phase amplitude fluctuation in the region is (see Fig.1):
\bea
\Delta \theta=\theta-\alpha={\sigma(x)\over \eta}-\alpha=arc tan{x\over l-y}~~,
\eea
where $\alpha$ , and $l$ are parameters shown in Fig.1. For the region $x\ll l$, $y\ll l$, we have:
\bea
\sigma(\vec x)\approx arc tan ({x\over l})\eta+\alpha\eta= \sigma_a arctan(\vec k\cdot \vec x)+ \Sigma,
\eea
where $\vec k=( 1/l,0,0)$ in the reference frame, $\sigma_a=\eta$, and $\Sigma=\alpha\eta$. Since
\bea
arctan(kx)\propto\int (e^{-|k'/k|}/k')e^{-ik'x}dk'~~,
\label{scale}
\eea
the perturbation is dominated by the normal modes with wavelengthes
\bea
(1/k')\gtrsim (1/k)=l~~,
\eea
which is scale dependent with a pivot scale $l$. As the universe expands, the wave lengthes of the perturbations expand as $1/k\propto a(t)$. The amplitude of the perturbation $\Delta \theta$ remains constant as far as the topological defect is inside the horizon. After the defect exits the horizon, the wave lengthes of the dominant modes stretch to super-horizon, and thus their amplitudes are frozen by causality according to Eq.(\ref{horizon}).

After inflation, the universe is reheated and is dominated by radiations. As the universe expands, the Hubble rate decreases. When $H\sim m_{\sigma}$, the Pseudo-Goldston field $\sigma$ rolls to its minimum so a cold condensate with a primordial fluctuation $\Delta \sigma=\Delta \theta \eta$ across the observable universe is created. The topological defect is well outside the observable universe. A domain wall could be created at $\theta=0$ \cite{Vilenkin:1982ks}. As we consider small fluctuations so $\Delta\sigma/\Sigma<1$, therefore $\Delta \theta<\alpha$ and thus the domain wall is outside the observable universe as well.

As the super-horizon wave vector $|\vec k|\ll H\lesssim m_\sigma$, the $\sigma$ particles behave as non-relativistic matter. The energy density $\rho_{\sigma}$ decreases proportional to $a^{-3}$. The dominant energy density $\rho$, which is radiation, decreases proportional to $a^{-4}$. Therefore $\rho_{\sigma}/\rho\propto a$, which implies that the energy density portion due to the $\sigma$ field grows before it decays to relativistic particles.

The $\sigma$ particles decay into radiation before the standard big bang nucleosynthesis (BBN) era in order to retain the success of the BBN. In this paper we consider only the standard particle decay scenario, so the decay rate of the $\sigma$ particle is:
\be
\Gamma={g\over 4\pi}{m_{\sigma}^3\over \eta^2}~~,
\ee
where $g$ is a model dependent factor which we take it order of one in the following discussions. Therefore we have
\be
\Gamma>H_B=\sqrt{{8\pi\over 3m_{pl}^2} {g_*\pi^2T_B^4\over30}}~~,
\ee
where $g_*\approx 10.8$ and $T_B\approx 1$MeV are the effective degrees of freedom of the radiation and the temperature at the BBN era respectively \cite{Olive:1999ij}. Let us assume $\eta=2T_c\sim 2.8\times10^{13}$GeV, then we have a weak boundary of the mass $m_{\sigma}\gtrsim 16.4$GeV. In addition, the energy density of the $\sigma$ particles at time $t$ just before their decay is:
\bea
\rho_{\sigma}\approx m_{\sigma}^2\eta^2[{a(t_1)\over a(t)}]^3=0.02{m_{\sigma}^5\over \eta}~~,
\label{sigma}
\eea
where $t_1\approx 1/m_{\sigma}$ is the time that the $\sigma$ field starts to oscillate and $t\approx 1/\Gamma$ is the lifetime of the $\sigma$ particles. The total energy density at time $t$ is:
\bea
\rho={g_*\pi^2T_B^4\over30}[{a(t_B)\over a(t)}]^4\approx1.2\times10^{35}{\rm GeV}^2{m_{\sigma}^6\over \eta^4}~~,
\label{tot}
\eea
where $t_B\sim H_B^{-1}\sim 1.5$s is the time of the BBN era. Combining Eq.(\ref{sigma}) and Eq.(\ref{tot}), gives the energy density ratio:
\be
{\rho_{\sigma}\over\rho}=1.7\times10^{-37}{\rm GeV}^{-2}{\eta^3\over  m_{\sigma}}~,
\label{mass}
\ee
after the $\sigma$ field decays. To preserve the homogeneity of the observable universe, $\rho_{\sigma}/\rho\ll1$, therefore, we have a stronger constraint: $m_{\sigma}\gtrsim 20$TeV.

\begin{figure}
\begin{center}
\includegraphics[width=0.9\textwidth]{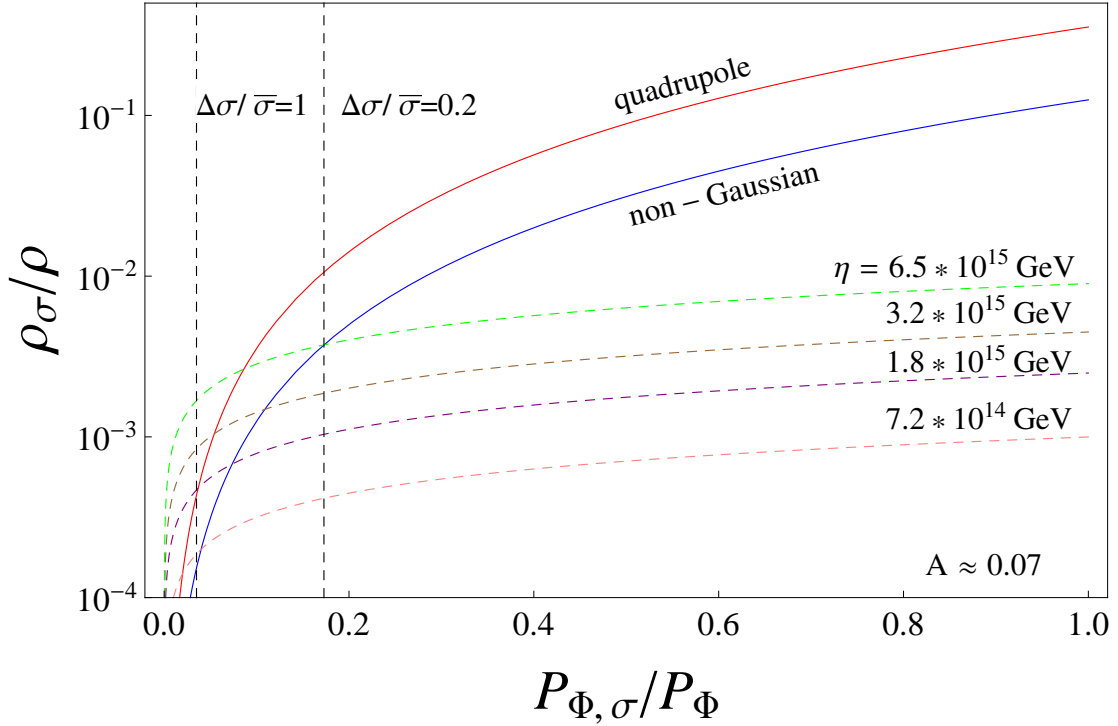}
\caption{The parameter space for the proposed scenario. The upper part of the parameter space is ruled out due to the CMB quadrupole constraint (red) and the lower part of the parameter space is ruled out due to the CMB non-Gaussian constraint (blue). The dashed lines are derived from Eq.(\ref{a3}) which relates symmetry breaking scale $\eta$, the density ratio, and the power spectrum ratio. The vertical line $\Delta\sigma/\bar\sigma$ indicates the magnitudes of small fluctuations. When all the cosmic factors are considered, the symmetry breaking scale $\eta\sim10^{15}$GeV, (see part V).}
\end{center}
\end{figure}
\section{The Power Asymmetry}
%{\itshape The Power Asymmetry:}
We now consider the perturbations in the CMB due to the $\sigma$ bosons. After the initial variation created by the topological defect, the fluctuation modes are inflated out of the horizon. The quantum effects give rise to an additional amplitude fluctuation $\delta\sigma(k)=(H_I/2\pi)|_{(k/a)=H_I}$ as the mass of the $\sigma$ field is negligible compared to the Hubble rate during inflation. The spectrum of the quantum created fluctuation is almost flat since $\dot H_I/H_I^2\ll 1$ and $|V_{\sigma\sigma}|\ll H_I^2$. Well after inflation and the era $H\sim m_{\sigma}$, the field starts to oscillate. The energy density of the $\sigma$ field is:
\be
\rho(\vec x)={1\over2}m_{\sigma}^2\bar\sigma^2(\vec x)~~,
\ee
where $\bar\sigma$ means the amplitude of the oscillation. Since $\bar\sigma\sim\eta \gg H_I$, the field perturbation is small and thus the energy density contrast is
\bea
{\delta\rho_{\sigma}(\vec x)\over \rho_{\sigma}(\vec x)}\approx2{\delta\sigma(\vec x)\over\bar \sigma(\vec x)}\sim {H_I\over \pi\bar \sigma(\vec x)}
\eea
where we have used the condition that the quantum fluctuation spectrum is flat. The resulting curvature perturbation is \cite{Lyth:2001nq}:
\be
P_{\Phi,\sigma}\approx ({\rho_{\sigma}\over\rho})^2({H_I\over\pi\bar \sigma})^2~,
\label{spectrum}
\ee
where $\rho_{\sigma}/\rho$ is defined at the epoch of the $\sigma$ decay. In addition, as $\rho_{\sigma}\ll\rho$ and $\rho_{\sigma}\propto\bar\sigma^2$, the curvature perturbation $P_{\Phi,\sigma}\propto \bar \sigma ^2$ and the perturbation of the curvature perturbation is $\delta P_{\Phi,\sigma}\propto 2\delta\sigma\bar \sigma$. The power spectrum asymmetry on large scales is then:
\bea
{\Delta P_{\Phi}\over P_{\Phi}}={\Delta P_{\Phi,\sigma}\over P_{\Phi,\sigma}}{ P_{\Phi,\sigma}\over P_{\Phi}}={2\Delta \sigma\over\bar \sigma}{P_{\Phi,\sigma}\over P_{\Phi}}\approx0.07~~,
\label{asymmetry}
\eea
where $\Delta\sigma/\bar \sigma=\sigma_aarctan(\vec k \cdot \vec x_{dec})/\Sigma\approx(\vec k\cdot \vec x_{dec})/\alpha$, and $x_{dec}$ denotes the decoupling in the comoving coordinate. Due to Eq.(\ref{scale}), the asymmetry is scale dependent:
\be
A(k)\propto {e^{-kl}\over k}~~.
\ee

In addition, the small inhomogeneities in the $\sigma$ field creates perturbations in the CMB through the Grishchuk-Zel'dovich effect \cite{L.Grishchuk}. The density fluctuation:
\be
\delta \rho=m^2[2\bar\sigma\delta\sigma+(\delta\sigma)^2]
\label{density}
\ee
generates a gauge-invariant curvature perturbation \cite{Bardeen:1983qw,Bassett:2005xm}, resulting in:
\bea
\Phi&\approx&-{1\over 5}{\rho_{\sigma}\over\rho}  {\delta \rho_\sigma\over \rho_{\sigma}}=-{1\over 5}{\rho_{\sigma}\over \rho}[2{\Delta \sigma\over \bar \sigma}+({\Delta \sigma\over \bar \sigma})^2]\nonumber\\
&=&-{1\over 5}{\rho_{\sigma}\over\rho}[2{\sigma_a\over \Sigma}(\vec k\cdot \vec x)+({\sigma_a\over \Sigma})^2(\vec k\cdot \vec x)^2+..]
\eea
This potential creates temperature fluctuations in the CMB by the (integrated) Sachs-Wolfe (SW) effect \cite{Sachs:1967er} and the Doppler effect. The resulting temperature fluctuation is \cite{Erickcek:2008jp}:
\be
{\Delta T\over T}(\hat n)=0.066\mu^2{\rho_{\sigma}\over\rho}({\sigma_a\over \Sigma})^2(\vec k\cdot \vec x_{de})^2+..~~,
\ee
where $\mu=\hat k\cdot \hat n$ and $\hat n$ is the observation direction. The dipole contribution vanishes because the SW effect and the ISW effect is canceled by the Doppler shift arising from the potential induced particle velocities, therefore the leading contribution term is quadratic and the observational constraint is from the quadrupole spherical-harmonic coefficient. Considering a 5$\sigma$ upper limit of the quadrupole temperature fluctuation $\sim5\times\sqrt{250(\mu K)^2}$ \cite{Efstathiou:2003tv}, the bound is: $({\rho_{\sigma}/\rho})({\sigma_a/\Sigma})^2(kx_{de})^2\lesssim 4.40\times10^{-4}$. This limit together with Eq.(\ref{asymmetry}) gives rise to an upper boundary in the Power spectrum ratio - Density ratio (PD) parameter space:
\bea
{\rho_{\sigma}\over\rho}\lesssim ({2\over 0.07})^2\times4.40\times10^{-4}\left(P_{\Phi,\sigma}\over P_{\Phi}\right)^{2}~~.
\label{a1}
\eea

As the $\sigma$ bosons produce non-Gaussian contributions to the density fluctuation by the $\delta\sigma^2$ term in Eq.(\ref{density}), there is an additional constraint. The non-Gaussian contribution is parameterized by $f_{NL}$ which is defined as: $\Phi=\Phi_{gauss}+f_{NL}\Phi_{gauss}^2$.
The resulting contributions is $f_{NL}={(5/ 4)}{(\rho/ \rho_{\sigma})}({P_{\Phi,\sigma}/ P_{\Phi}})^2$ \cite{Verde:1999ij,Malik:2006pm,Ichikawa:2008iq}.
The Planck mission finds that $f_{NL}\lesssim 0.01\%\times10^5$ \cite{Ade:2015ava}, thus the lower boundary in the PD space is:
\be
{1\over 8}\left({P_{\Phi,\sigma}\over P_{\Phi}}\right)^2\lesssim {\rho_{\sigma}\over \rho}~~.
\label{a2}
\ee

Lastly, the magnitude of the primordial power spectrum is of the order of $P_{\Phi}\sim 1.5\times10^{-9}$ \cite{Ade:2015lrj}. Giving a particular symmetry breaking scale $\eta$, Eq.(\ref{spectrum}) yields:
\be
{\rho_{\sigma}\over \rho}\approx\sqrt{\left({P_{\Phi,\sigma}\over P_{\Phi}}\right)\times1.5\times10^{-9}}\left({\eta\pi\over H_I}\right)~~.
\label{a3}
\ee
Therefore, we put the constraints from Eq.(\ref{a1}), Eq.(\ref{a2}), and the relationships due to Eq.(\ref{a3}) into the PD space, see Fig.(2). Notice that the field fluctuation across the observed universe $\Delta\sigma/\bar \sigma<1$ for small fluctuation. It gives a left boundary of the $P_{\Phi,\sigma}/P_{\Phi}$, thus $\eta\gtrsim 7.2\times10^{14}$GeV. In addition, Eq.(\ref{mass}) yields a particular mass for a given $\eta$ and $\rho_{\sigma}/\rho$. As the $\sigma$ bosons start to oscillate after inflation, $m_{\sigma}\ll H_I$, we have an upper boundary $\eta\lesssim 6.5\times10^{15}$GeV. Therefore in a scenario that an inflaton plus a Pseudo-Goldstone, we have $\eta\sim 10^{15}$GeV and $m\sim 10^{12}$GeV. The explicit symmetry breaking potential of the $\sigma$ field therefore has a form $V(\sigma)\sim(\eta^2/m_{pl})\sigma^2+{\cal O}(\sigma^2)$.

\section{Summary and Discussion}
%{\itshape Summary and Discussion:}
Observations of Cosmic Microwave Background anisotropy suggest that there are "anomalies" in the large scale structures of the CMB. The hemispherical power asymmetry may indicate a possible pre-inflationary origin. In this paper, we have suggested a scenario that a string-like topological defect was created by spontaneous symmetry breaking prior to inflation. This defect then generated a spacial perturbation in a boson field. The amplitude of this perturbation is protected by topology before the defect exits horizon and is frozen by causality after the defect is inflated out of the horizon. Well after inflation, the boson field started to oscillate with the initial perturbation across the observable universe. The created bosons then decayed into radiations before the BBN era.

We have shown that the density perturbation of the bosons can generate the required power asymmetry without producing unwanted multi-poles and non-Gaussianity fluctuations in the CMB. While we consider here a complex scalar field with a U(1) symmetry potential plus a mass term, the underlying mechanism can be general.

\section*{ACKNOWLEDGMENTS}
%{\itshape Acknowledgments:}
We would like to thank JianWei Cui for helpful discussions. This work is partially supported by the Natural Science Foundation of China under Grant Number 11305066.

\end{document}